\documentclass[twoside,twocolumn,9pt]{article}
\usepackage{extsizes}
\usepackage[super,sort&compress,comma]{natbib} 
\usepackage[version=3]{mhchem}
\usepackage[left=1.5cm, right=1.5cm, top=1.785cm, bottom=2.0cm]{geometry}
\usepackage{balance}
\usepackage{times,mathptmx}
\usepackage{sectsty}
\usepackage{graphicx} 
\usepackage{lastpage}
\usepackage[format=plain,justification=justified,singlelinecheck=false,font={stretch=1.125,small,sf},labelfont=bf,labelsep=space]{caption}
\usepackage{float}
\usepackage{fancyhdr}
\usepackage{fnpos}
\usepackage[english]{babel}
\addto{\captionsenglish}{%
  
}
\usepackage{array}
\usepackage{droidsans}
\usepackage{charter}
\usepackage[T1]{fontenc}
\usepackage[usenames,dvipsnames]{xcolor}
\usepackage{setspace}
\usepackage[compact]{titlesec}
\usepackage{hyperref}

\usepackage{epstopdf}

\definecolor{cream}{RGB}{222,217,201}

\begin{document}

\pagestyle{fancy}
\thispagestyle{plain}
\fancypagestyle{plain}{

\fancyhead[C]{\includegraphics[width=18.5cm]{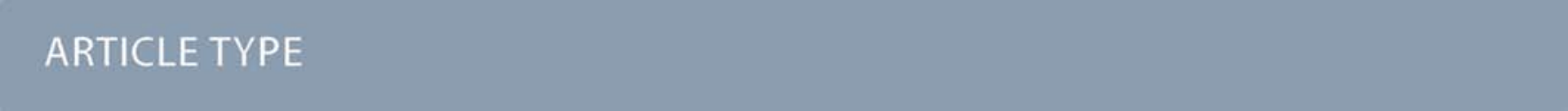}}
\fancyhead[L]{\hspace{0cm}\vspace{1.5cm}\includegraphics[height=30pt]{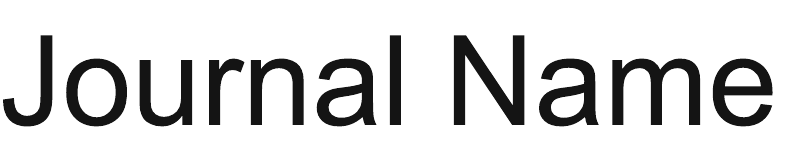}}
\fancyhead[R]{\hspace{0cm}\vspace{1.7cm}\includegraphics[height=55pt]{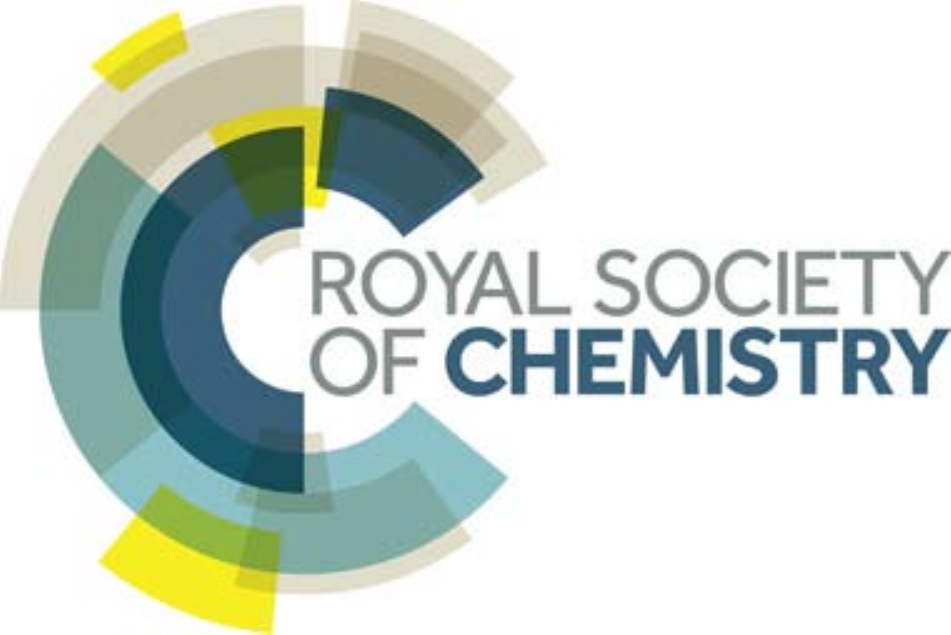}}
\renewcommand{\headrulewidth}{0pt}
}

\makeFNbottom
\makeatletter
\renewcommand\LARGE{\@setfontsize\LARGE{15pt}{17}}
\renewcommand\Large{\@setfontsize\Large{12pt}{14}}
\renewcommand\large{\@setfontsize\large{10pt}{12}}
\renewcommand\footnotesize{\@setfontsize\footnotesize{7pt}{10}}
\makeatother

\renewcommand{\thefootnote}{\fnsymbol{footnote}}
\renewcommand\footnoterule{\vspace*{1pt}%
\color{cream}\hrule width 3.5in height 0.4pt \color{black}\vspace*{5pt}} 
\setcounter{secnumdepth}{5}

\makeatletter 
\renewcommand\@biblabel[1]{#1}            
\renewcommand\@makefntext[1]%
{\noindent\makebox[0pt][r]{\@thefnmark\,}#1}
\makeatother 
\renewcommand{\figurename}{\small{Fig.}~}
\sectionfont{\sffamily\Large}
\subsectionfont{\normalsize}
\subsubsectionfont{\bf}
\setstretch{1.125} 
\setlength{\skip\footins}{0.8cm}
\setlength{\footnotesep}{0.25cm}
\setlength{\jot}{10pt}
\titlespacing*{\section}{0pt}{4pt}{4pt}
\titlespacing*{\subsection}{0pt}{15pt}{1pt}

\fancyfoot{}
\fancyfoot[LO,RE]{\vspace{-7.1pt}\includegraphics[height=9pt]{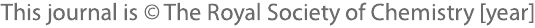}}
\fancyfoot[CO]{\vspace{-7.1pt}\hspace{13.2cm}\includegraphics{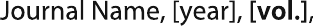}}
\fancyfoot[CE]{\vspace{-7.2pt}\hspace{-14.2cm}\includegraphics{head_foot/RF}}
\fancyfoot[RO]{\footnotesize{\sffamily{1--\pageref{LastPage} ~\textbar  \hspace{2pt}\thepage}}}
\fancyfoot[LE]{\footnotesize{\sffamily{\thepage~\textbar\hspace{3.45cm} 1--\pageref{LastPage}}}}
\fancyhead{}
\renewcommand{\headrulewidth}{0pt} 
\renewcommand{\footrulewidth}{0pt}
\setlength{\arrayrulewidth}{1pt}
\setlength{\columnsep}{6.5mm}
\setlength\bibsep{1pt}

\makeatletter 
\newlength{\figrulesep} 
\setlength{\figrulesep}{0.5\textfloatsep} 

\newcommand{\topfigrule}{\vspace*{-1pt}%
\noindent{\color{cream}\rule[-\figrulesep]{\columnwidth}{1.5pt}} }

\newcommand{\botfigrule}{\vspace*{-2pt}%
\noindent{\color{cream}\rule[\figrulesep]{\columnwidth}{1.5pt}} }

\newcommand{\dblfigrule}{\vspace*{-1pt}%
\noindent{\color{cream}\rule[-\figrulesep]{\textwidth}{1.5pt}} }

\makeatother

\twocolumn[
  \begin{@twocolumnfalse}
\vspace{3cm}
\sffamily
\begin{tabular}{m{4.5cm} p{13.5cm} }

\includegraphics{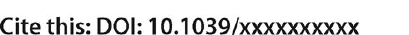} & \noindent\LARGE{\textbf{Surface forces generated by the action of electric fields across liquid films}} \\\vspace{0.3cm} & \vspace{0.3cm} \\

 & \noindent\large{Carla Sofia Perez-Martinez and Susan Perkin$^{\ast}$} \\

\includegraphics{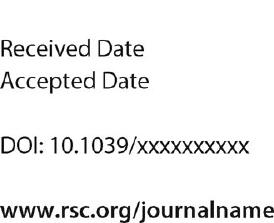} & \noindent\normalsize{We explore the force generation and surface interactions arising when electric fields are applied across fluid films. Using a surface force balance (SFB) we measure directly the force between two electrodes in crossed-cylinder geometry across dielectric and electrolytic fluids. In the case of dielectric films the field between the electrodes exerts a force which can be well explained using classic expressions and with no fitting parameters. However when the electrodes are separated by a film of electrolyte, an alternating electric field induces a force which diverges substantially from the calculated static response of the electrolyte. The magnitude of the force is larger than predicted, and the interaction can switch from attractive to repulsive. Furthermore, the approach to steady state in electrolyte takes place over $10^2$ -- $10^3$~s which is very slow compared to both the charging and viscous timescales of the system. The non-trivial electrolyte response in AC electric fields, measured here directly, is likely to underlie several recent reports of unexpected and bifurcating forces driving colloids in AC fields. Our measurements suggest ways to control colloidal and soft matter using electric fields, as well as providing a direct measure of the length- and time-scales relevant in AC electrochemical and electrokinetic systems. }
\end{tabular}
 \end{@twocolumnfalse} \vspace{0.6cm}
  ]

\renewcommand*\rmdefault{bch}\normalfont\upshape
\rmfamily
\section*{}
\vspace{-1cm}

\footnotetext{\textit{~Department of Chemistry, Physical and Theoretical Chemistry Laboratory, University of Oxford, Oxford OX1 3QZ, United Kingdom; E-mail: susan.perkin@chem.ox.ac.uk}}
\footnotetext{\dag~Electronic Supplementary Information (ESI) available: Discussion of electrochemical effects and Joule heating; tables of parameters; curve fitting. See DOI: 10.1039/b000000x/}

\section{Introduction}

Electric fields act across liquid or soft films in many technological and natural contexts, such as in liquid crystal displays\cite{DeGennes1995}, touch screens\cite{Ayyildiz2018}, batteries\cite{Bokris1995}, cell membranes\cite{HodgkinHuxley1952}, and in the shocking party-tricks of electric fish\cite{Faraday1839,HodgkinHuxley1952,Catania2014}. 
 In purely dielectric materials the strength of the electric field, $E$, scales with the potential difference and inversely with the film thickness: $E = V/D$. Importantly, for films of thickness $D \sim 10~\mu{\rm{m}}$ and below, only a modest voltage need be applied to achieve an electric field energy density comparable to the thermal energy per molecule, \textit{i.e.} $\epsilon E^2 \approx k_{\rm B}T/l_m^3$ (with $\epsilon$ the permittivity of the material, $k_{\rm B}$ Boltzmann's constant, $T$ the temperature and $l_m$ a molecular dimension). That is to say, the electric field injects energy at a level sufficient to compete with molecular interactions in liquids and soft matter.  In this way electric fields can be used to control dynamic and structural transitions in films of soft matter, remotely and reversibly.

\begin{figure}
\centering
\includegraphics{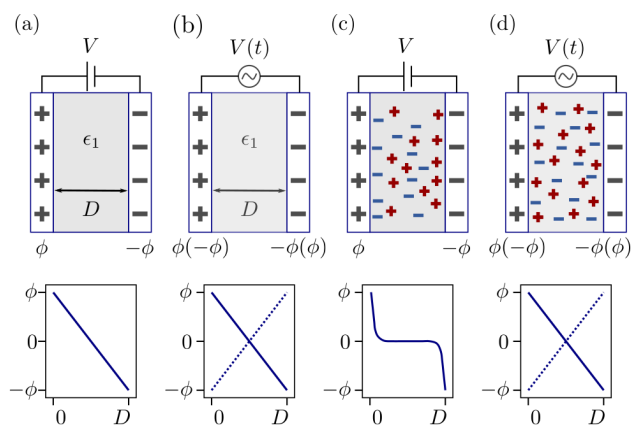}
\caption{Dielectric and electrolytic fluids under static and oscillating electric fields; schematic diagrams representing various scenarios discussed in the introduction. (a)~Static field applied across a dielectric film of thickness $D$ (top) and the linear drop in potential, $\phi(x)$ (bottom); (b)~Time-varying (AC) field (top) applied to a dielectric film and the limiting gradients of potential (bottom); (c)~Static field applied across electrolyte (assuming no Faradic process) showing electrical double layer formation (top) and the concomitant screening of the potential so that the electric field (potential gradient) is high near the surfaces and approaches zero at the midplane, which is the case when $D >> \lambda_S$ (bottom); (d)~AC field applied across an electrolyte with $1/\tau_{AC} >> 1/\tau_{DL}$ such that the double layers cannot rearrange sufficiently rapidly to screen the surface charges (top), in which case it might be expected that $\phi(x)$ is again linear (bottom), but see discussion.}
\label{Dielec}
\end{figure}

The electric field acts not only on the fluid medium but also introduces a \textit{surface force} between the oppositely polarised electrodes. This surface force is a convenient measurable quantity which allows direct insight into the relaxation times\cite{Tivony2018}, dynamic transitions\cite{Drummond2012,Kristiansen2015}, length-scales and other properties of soft films. Other than these pioneering works there have been rather few reports of experimental measurement of field-induced surface forces; in part this is due to the difficulty of setting up sufficiently well-controlled experiments to obtain unambiguous quantitative information. It is this field-induced surface force which is the focus of our present study, where we present measurements in air and an oil (as controls) then in ionic liquid electrolytes. First, we provide some introductory remarks on the force acting across dielectric and electrolytic media when an electric field is applied perpendicular to the film, and summarise some less-understood observations when AC fields are applied in colloidal and electrolytic systems. 

\subsection*{Field-induced forces across dielectric fluids}
When a voltage is applied between electrodes bounding a purely dielectric film (permittivity $\epsilon_1$ and thickness $D$) the electrostatic potential decays linearly across the fluid; Figure~\ref{Dielec}(a). As long as no current passes this is a simple capacitor, and the stored electrical energy  $U^{ES}$ is given by 
 
 \begin{equation} 
U^{ES} =  \tfrac{1}{2}CV^2
\label{Ues}
\end{equation}
where $C = \epsilon_1/D$ is the capacitance per unit area and $V = (\phi_1 - \phi_2) = \Delta \phi$ is the voltage drop between the surfaces. The electrostatic potential in the medium between the electrodes, $\phi(x)$, varies linearly from $\phi_1=\phi$ to $\phi_2=-\phi$ between $x=0$ and $x=D$. The excess pressure between the plates can be calculated by considering the surface charge density, $\sigma$, arising from the applied potential: 
\begin{equation} 
\Pi^{\rm dielectric} =  -\tfrac{1}{2}\sigma E = -\tfrac{1}{2}\epsilon_1 E^2 = -\tfrac{1}{2}\epsilon_1 \left(\frac{V}{D}\right)^2.
\label{F}
\end{equation}
And so it is that capacitor plates attract one another with magnitude determined by the square of the applied voltage, and increasing linearly with the dielectric constant of the medium. When the potential applied between the electrodes varies in time sinusoidally between $+V$ and $-V$ at frequency $\nu_{\rm{AC}} = 1/{\tau_{AC}}$ (Figure~\ref{Dielec}(b)) then the excess pressure between the surfaces maintains a negative (attractive) value because the square of the potential gradient is always positive. The magnitude of the attractive pressure varies during the AC cycle, i.e. $\Pi(t) = -1/2\epsilon E(t)^2$, and the capacitance and surface charge magnitudes may (depending on $\nu_{\rm{AC}}$) also vary due to the frequency dependence of the dielectric function\cite{Kramer2003}. 

\subsection*{Force across electrolyte in a static field}
The situation becomes more complex and interesting when the fluid contains dipolar and charged species: the electric field exerts a force on the charges and so drives fluxes at the onset of the field and concentration gradients at steady-state,  with the details depending delicately on the electrolyte properties and the interplay between the natural relaxation times of the fluid and the applied field frequency, $\nu_{\rm{AC}}$\cite{Bazant2004}.  A static electric field ($\nu_{\rm{AC}} = 0$) will drive the formation of an electrical double layer in the electrolyte near the electrodes. At low ion density, when the ion distributions and charge density can be well described by the Poisson-Boltzmann equation, this results in exponential decay of $\phi(x)$, and of $E = -\left(\frac{{\rm d}\phi_m}{{\rm d}x}\right)$, with a decay length called the screening length, $\lambda_S$ (which is equal to the Debye screening length at asymptotically low concentration), as is well documented\cite{IsraelachviliB} (see Figure~\ref{Dielec}(c)).  Of particular relevance to the present work is the excess pressure between the electrodes, and for an electrolyte at equilibrium in a static field this has two contributions: the excess osmotic pressure of counterions in the electrical double layers, and the electric field contribution which is proportional to the gradient of potential at the midplane\cite{BenYaakov2007,Silbert2012}:
\begin{equation} 
\Pi^{\rm electrolyte}=2  k_{\rm B}T\rho_\infty \left({ \rm cosh}(e\phi_m /k_{\rm B}T)-1\right) -\frac{\epsilon_1}{2}\left(\frac{{\rm d}\phi_m}{{\rm d}x}\right)^2
\label{PiEDL}
\end{equation}
where $\rho_\infty$ is the bulk ion density and $\phi_m$ is the electrostatic potential at the midplane ($x = D/2$). In symmetric systems ($\phi_1 = \phi_2$) the electric field term vanishes and the pressure is simply the osmotic pressure. However whenever the surfaces are asymmetric ($\phi_1 \neq \phi_2$) there is an electric field contribution. The details of how this surface interaction manifests itself under various conditions is quite complex and has been studied theoretically and experimentally for many years \cite{Parsegian1972,Kampf2009,BenYaakov2007,Silbert2012,Majee2018}. For the present work two important aspects are most relevant: first, when the surfaces are separated by distances $D \gg \lambda_S$ both the osmotic and electric field contributions approach zero and so, in contrast to the dielectric case, we expect to measure no force in a static field at large surface separations. The second important aspect, which we consider next, concerns the behaviour during the transient when the field is switched on, or when the field is time-varying. 

\subsection*{Electrolyte in a transient or oscillating field}
Electrokinetic theories have been constructed to interpret and predict the dynamics of charged particles under influence of static and oscillating fields \cite{OBrien1978,Delacey1981,Anderson1989,Bazant2010}. The focus of these works has primarily been the motion of charged colloids in steady electric fields (electrophoresis)\cite{Anderson1989} and, later, the generation of fluid flows under AC pumping (AC electroosmosis, ACEO) \cite{Ramos1999,Ajdari2000}.  The general starting point is to couple Poisson's equation for the charge-density dependence of electrostatic potential with the Nernst-Planck equation describing fluid flow under the influence of concentration gradients and electric field. Several strong assumptions are typically made: low ion density, identical diffusivities for cation and anion ($\mathcal{D}_+ = \mathcal{D}_- =\mathcal{D}$), no inertial effects, and small electric field magnitude. 

Careful work has also been dedicated to the question of the electrolyte response during transients; either when a sudden DC field is applied or with AC field. For example Bazant \textit{et al.} have provided an analysis of the evolution of potential and ion density across an electrolyte as a function of time after an applied DC field \cite{Bazant2004}. The timescale expected to dominate the transient response is the double-layer charging time and scales as $\tau_{c} \sim {\lambda{_S}D}/{\mathcal{D}}$ for parallel plates separated by distance $D$; for a cylindrical pore the transmission line model of De Levie\cite{DeLevie1963,Mirzadeh2014} predicts that $\tau_{c}$ should be modified by a geometrical factor. Recent experiments verify this does indeed capture the experimental double layer charging time in confined dilute electrolyte\cite{Tivony2018}.  Further analyses addressed the situation of high voltage or high ionic strength when the typical assumptions break down\cite{HojgaardOlesen2010,Storey2012}.

Despite this progress a number of recent experimental observations of the influence of AC fields on particles in electrolyte remain difficult to interpret, even in qualitative terms.  Oscillating electric fields imposed across particles immersed in electrolyte induce strong and long-ranged particle-electrode and particle-particle interactions\cite{Trau1996,Yeh1997}, and the direction of the force can be altered from attractive to repulsive depending on the experimental conditions\cite{Kim2002,Hoggard2007,Hoggard2008,Wirth2013,Woehl2014,Woehl2015,Bukosky2015}. Furthermore, XPS measurements have shown that voltage transients can be measured up to millimetres away from an electrified interface in ionic liquids \cite{Camci2016}. Recent theoretical studies also paint an intricate picture of non-equilibrium effects in electrolytes under AC fields leading to long-range concentration gradients\cite{Khair2018,Hashemi2018,HojgaardOlesen2010,Schnitzer2014,Stout2015} and in particular recent works highlight the strong relevance of asymmetry in the cation and anion diffusivity, both during the transient after switching the field \cite{Khair2018} and in AC fields \cite{ Hashemi2018}. 

The concomitant \textit{force} between the electrode surfaces resulting from the ion motion has been less considered, although this should reveal the relevant timescales during transients and the nature of electrolyte screening under different situations. For example, a na\"ive expectation might be  that at low AC frequency the electrolyte screens the surface charges and no force acts between the two electrodes (as in Figure~\ref{Dielec}(c)), whereas at very high AC frequency the electrolyte ions cannot translate during one AC cycle and so the electrolyte behaves as a dielectric medium; Figure~\ref{Dielec}(d). Clearly much interest and complexity lies between these limiting cases -- \textit{i.e.} under varying $\nu_{\rm{AC}}$ but also varying ion density, voltage, and electrode geometry.

\subsection*{Prospectus}
In this work we report our explorations of the force acting between electrodes across dielectric and electrolyte films. We used a surface force balance (SFB) to measure the interaction force between the electrode surfaces as a function of their separation distance and applied DC or AC voltage, across several test fluids (dry gas, a dielectric liquid, and two ionic liquids). Both the transient forces after switching the applied field and the steady-state interaction force between the surfaces were measured, and we compare the results to calculations of the equilibrium and transient forces in the system. We find that measurements with dielectric fluids match closely to the calculated behaviour, during the transient and at steady-state, with no fitting parameters. This close matching of theory and experiment for the dielectric materials is an important prerequisite in ensuring the later measurements with electrolytes are robust, and so we discuss in detail these control experiments.  Finally, we then report our measurements with ionic liquids between the electrodes: the forces diverge dramatically from those measured with non-electrolytes, and do not appear to be tamed by existing electrokinetic theories. 

The manuscript is arranged as follows: In section \ref{methods} we describe the experimental setup; in section \ref{dielectrics} we summarise some basic results for the expected force between capacitor plates across dielectrics arranged in the geometry of our experiment; then in section \ref{results} we describe and interpret our measurements, first with dielectric media (air then the non-polar liquid polydimethylsiloxane), and finally with ionic liquids.  


\section{Experimental Setup} \label{methods}
\subsection{Surface Force Balance modified for application of electric fields perpendicular to the liquid film}

\begin{figure*}
\includegraphics{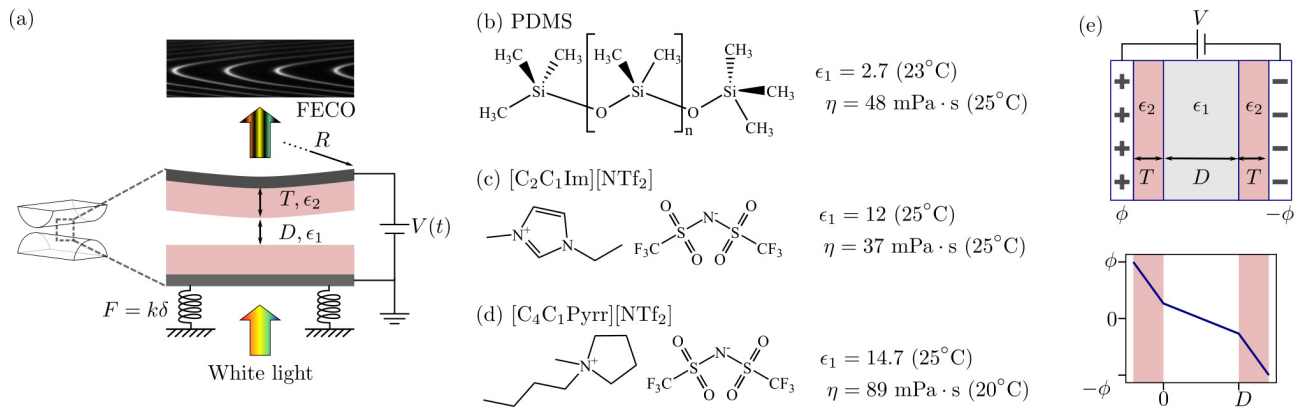}
\caption{ (a) SFB experimental geometry, showing crossed-cylinder geometry of the lenses used in the experiment and the detail of the layers of the setup. The liquid film of thickness $D$ is separated from each silver electrode by a mica spacer layer of thickness $T$. The crossed-cylinder experimental geometry has an effective radius of curvature $R$. The normal force $F$ arising between the surfaces is measured using the deflection $\delta$ of a spring of constant $k$. White light is incident on the three-layer interferometer, and the resulting pattern of fringes of equal chromatic order (FECO) is used to determine $D$, $F$ and $R$. (b) Polydimethylsiloxane, a viscous oil used as a control in the electric field experiments. Properties from Ref.~\cite{Hakim1977}. (c) 1-ethyl-3-methylimidazolium bis(trifluoromethylsulfonyl)imide, properties from Ref.~\cite{Huang2011,Yu2012} (d) 1-butyl-1-methylpyrrolidinium bis(trifluoromethylsulfonyl)imide, properties from Ref.~\cite{Huang2011,Yu2012} (e) Static field applied across a 3-layer capacitor, with similar dielectric layers on each electrode (of thickness $T$ and dielectric constant $\epsilon_2$ and a medium between of thickness $D$ (top) and potential drop across the three layers for $\epsilon_2 > \epsilon_1$ (bottom)}
\label{Figure:Experimental}
\end{figure*}

Experiments were performed using a surface force balance (SFB) wherein white light interferometry is used to determine the separation distance, $D$, and normal force, $F$, between two surfaces separated by a thin liquid film. In the present work electric fields were applied normal to the film: the (semi-transparent) mirrors used for interferometry acted simultaneously as electrodes. In the configuration used for most experiments, sketched in Figure~\ref{Figure:Experimental}(a), atomically smooth mica sheets of precisely equal thickness $T$ lie between each mirror/electrode and the liquid film. The mica sheets act to block Faradic processes at the electrodes, and also facilitate our measurement of film thickness down to molecular dimensions which is helpful in ascertaining cleanliness of the experiments. To ensure no mica-specific effects, some experiments were conducted with no mica layer, \textit{i.e.} $T=0$. The key aspects of the experiment have been described before\cite{Perkin2006a} and we refer to an article describing several recent modifications\cite{Lhermerout2018}. The procedure for applying perpendicular electric fields is similar to that described previously by Drummond \cite{Drummond2012}. Details particular to the present work are as follows. 

Mica was cleaved along the crystal planes resulting in large pieces (area $\approx~1~\rm{cm}^2$) with no steps in the crystal on either side and thickness $T \approx 1-3~\rm{\mu m}$ (then measured precisely in each experiment). These were backsilvered ($\approx 30-40$~nm Ag) then glued, silver side down, on cylindrical silica lenses (radius of curvature $R$~$\approx$~1~cm; measured precisely). The lenses were then mounted in a crossed-cylinder configuration, and a liquid droplet injected between the mica surfaces. The silver-mica-liquid-mica-silver stack forms an interferometric cavity: white light incident normal to the film emerges as an interference pattern (called Fringes of Equal Chromatic Order, FECO) from which the geometry and separation can be precisely determined (precision 0.1~nm).  The bottom lens is mounted on a spring (of pre-calibrated stiffness~$k$), whose deflection perpendicular to the cylinder axis, $\delta$, is also determined from the interference spectrum. The force is then $F=k\delta$.

Electric fields were applied perpendicular to the liquid film using the silver mirrors simultaneously as electrodes\cite{Drummond2012}.  Importantly, the crossed-cylinder geometry of our experiment means that the electric field also has a (weak) lateral gradient due to the variation in $D$ away from the point of closest approach. A controlled voltage was applied to the electrode on the top lens, relative to the electrode on the bottom lens, which was held at ground potential.  A thin Au or Cu wire was contacted to each silver surface through a small piece of conductive glue (made from Shell EPON 1004 and graphite). The Au wires are provided by Goodfellow, 0.125 mm Au diameter, with a PTFE coating of 0.016~mm thickness. The Cu wires (Goodfellow) are 0.05 mm diameter, with a polyesterimide coating of 0.002 mm thickness. We verified that the wires did not affect the bending of the spring where the lens is mounted through measurements of the spring constant with and without wiring in place. The measured spring constant with wiring is within the error of the measured $k$ without wires. A Keysight 33500B is used to apply the voltages to the electrodes. The bottom electrode is contacted to the electrical ground of the chamber through the signal generator, and the bias is always applied to the top electrode. Prior to each experiment with liquid, the electrode connections were tested and calibrations carried out in air to check for the expected attractive forces as in section~\ref{Results-air}. 

In a typical experiment the surfaces were moved (using a mechanical drive) to a chosen starting separation, $D_0$, and the distance was recorded for at least 30 seconds prior to applying a voltage at time $t=0$. The system response was recorded over time until reaching steady state; once $D(t)$ is stable, the voltage was set back to zero and the relaxation was again recorded. Spectrograms were recorded at a rate of 10 frames per second throughout each experimental run. The resulting $D(t)$ traces usually contained small superimposed drift due to the very long timescale of the measurements reported here; the traces presented in this paper have been corrected to subtract a linear drift when necessary. 

For some of the experiments the electrode surface was in direct contact with the liquid (no mica spacer layers; $T = 0$). In this case gold was used as the mirror/electrode material. A layer of chromium (2 nm) was thermally evaporated directly onto the fused silica lenses, and then a layer of gold (40 nm thickness) evaporated on top of the chromium. Before and after evaporation, the lenses are cleaned in solvents (toluene, choloroform and ethanol), rinsed in deionised water, treated with piranha solution (3 H$_2$SO$_4$:1 H$_2$O$_2$), and rinsed in deionised water and ethanol. The gold surfaces are also contacted using Au or Ag wires, held in place by the same conductive glue described above, and all connections and procedures for applying the electric field and recording the system response were as described for the mica setup.

\subsection{Materials}

Mica (ruby muscovite) was of optical grade (S\&J Trading). Figure~\ref{Figure:Experimental} shows the chemical structures of the liquids used, along with their viscosities and their extrapolated zero-frequency dielectric constants. Polydimethylsiloxane (PDMS) was used as a control non-polar liquid of similar viscosity to the ionic liquids (PMX200 50 cSt melt from Xiameter, used as received). The viscosity, from the manufacturer, is $\eta$=0.048~Pa$\cdot$s, and the dielectric constant is 2.7 \cite{Hakim1977}. The two ionic liquids used were 1-ethyl-3-methylimidazolium bis(trifluoromethylsulfonyl)imide, [C$_2$C$_1$Im][NTf$_2$], and 1-butyl-1-methylpyrrolidinium bis(trifluoromethylsulfonyl)imide, [C$_4$C$_1$Pyrr][NTf$_2$]. Both ionic liquids are 99\% purity (Iolitec). Prior to experiments, the ionic liquid is dried in vacuo ($< 5 \times 10^{-3}$~mbar) at 80~$^\circ$C for at least 12 hours before injecting it in the SFB. Once the liquid was injected between the SFB lenses, the chamber was sealed and purged with N$_2$. Fresh P$_2$O$_5$ powder was present in the chamber as an additional desiccant. 


\section{Crossed-cylinder capacitor forces and the mechanical response of the SFB} \label{dielectrics}
Before describing and interpreting our experiments we note down the general form of the electrical force between electrodes in the crossed-cylinder geometry of our experiment, and the resulting prediction for the deflection of the spring during a transient and in an AC field. The transient must account for the mechanical response of the experiment including the viscous force acting during motion of the surfaces. Understanding the AC scenario is necessary as a backdrop for electrolyte measurements and comparison to results of AC electro-phoretic / osmotic experiments,  impedance or dielectric spectroscopy, and for avoidance of Faradic reactions. 

\subsection*{Dielectric film with dielectric spacers (3-layer capacitor)}
In many experiments (and applications) it is necessary to introduce a dielectric spacer layer between the electrode and the film; this can be to block Faradic processes at the electrode surface or to alter the surface chemistry or optical properties in some desirable way.  The parameters of this 3-dielectric-layers system and the corresponding potential drop between the electrodes are shown in Figure~\ref{Figure:Experimental}(e).  The field passes through each of the dielectric layers in series, and the system can be treated as a three-layer capacitor. If the two spacer layers are identical in material and thickness then the capacitance becomes  
\begin{equation} 
\frac{1}{C_{\rm TOT}} = \frac{1}{C_1}+\frac{2}{C_2} = \frac{D}{\epsilon_1} +\frac{2T}{\epsilon_2}
\label{Ctot}
\end{equation}
where $C_1$ and $C_2$ are the capacitances of the film and spacers respectively, $\epsilon_1$ is the permittivity of the film and $\epsilon_2$ is the permittivity of the spacer. And so we have for the energy stored in the 3-layer capacitor, from Equations ~\ref{Ues} and ~\ref{Ctot}:
\begin{equation} 
U^{ES} =  \frac{V^2}{2(2T/\epsilon_2+D/\epsilon_1)}
\label{Eqn:Ues}
\end{equation}

In the experiments reported here the spacer layers are mica sheets (permittivity $\epsilon_2 = 8 \epsilon_0$ ). 

\subsection*{Dynamics: Equation of Motion} \label{DoM}

We now consider the time-evolution of the surface force and resulting film thickness, $D(t)$, under the influence of the (time-varying) electric field and mechanical forces of the system. In the case of a dielectric fluid, it is possible to predict $D(t)$ when the electrodes are subjected to a potential difference $V(t)$. Let us define $D(t=0)=D_0$ as the initial film thickness, and recall $\delta$ is the displacement of the spring, so that $D(t)=D_0-\delta$.

A simple force balance for a mass $M$ on a spring (of constant $k$), subjected to an electric field force $F_e$ and a drag force $F_{drag}$, gives
\begin{equation}
M\frac{d^2\delta}{dt^2}+ k\delta = F_e + F_{drag}
\label{Eqn:EquationofMotion}
\end{equation}

The drag force due to drainage of a viscous fluid during approach of crossed-cylinders (or a sphere approaching a flat wall), as appropriate for the SFB experimental setup, has been derived previously by Chan and Horn\cite{Chan1985}, and is given by

\begin{equation}
F_{drag}=\frac{6\pi\eta R^2}{D}\frac{dD}{dt}=-\frac{6\pi\eta R^2}{(D_0-\delta)}\frac{d\delta}{dt}
\label{Eqn:Drag}
\end{equation}

In general, the drag in Equation~\ref{Eqn:Drag} should include a negative slip length~$b$, so that the drag scales as $1/(D-b)$ instead of $1/D$. The $b$ correction accounts for the stopped layers of molecules adjacent to the surfaces that do not squeeze out during an approach. For the ionic liquids in question, the value of $b$ is of the order of a few nanometers \cite{Lhermerout2018}, while we verified the value of $b$ to also be within 5~nm for PDMS using the methodology from \cite{Lhermerout2018}. In our experiments, relatively large surface separations of $D>400$~nm are used; at these separations, it is safe to neglect the $b$ correction.

The force due to the electric field in the curved SFB geometry can be expressed using Derjaguin's approximation (which applies for $R>>D$, as is the case in our setup), 

\begin{equation}
F_e=2\pi RU^{ES}=\frac{\pi R V^2}{(2T/\epsilon_2+D/\epsilon_1)}=\frac{\pi R V^2}{(2T/\epsilon_2+(D_0-\delta)/\epsilon_1)}
\label{Eqn:Fe}
\end{equation}

where $U^{ES}$ is the electrostatic energy as in Equation \ref{Eqn:Ues}. Note that in general $V$ can be a time-varying function. Let us consider a sinusoidally oscillating voltage of the form $V(t)=V_0\cos{\omega t}$ (with $\omega=2\pi\nu_{\rm{AC}}$).  The DC case is obtained when $\omega=0$.

Equation \ref{Eqn:EquationofMotion} is a second-order, non-linear differential equation, but it is possible to obtain an approximate analytical solution if the displacement is small compared to the separation of the surfaces, this is, $\delta<<D$. In that case, we can approximate  $D_0-\delta \approx D_0$. We define

\begin{equation}
\gamma=\frac{6\pi \eta R^2}{D_0}
\end{equation}

and 
\begin{equation}
F_0= \frac{\pi R V_0^2}{2T/\epsilon_2+D_0/\epsilon_1}
\label{Eqn:F0vsV}
\end{equation}

so that equation \ref{Eqn:EquationofMotion} can be rewritten as 

\begin{equation}
M\frac{d^2\delta}{dt^2}+\gamma \frac{d\delta}{dt} + k\delta = F_0 \cos^2{\omega t} 
\end{equation}

The solution is of the form

\begin{equation}
\delta(t)=A_1e^{-t/\tau_1}  + A_2e^{-t/\tau_2} + \frac{F_0}{2k} + \frac{F_0 \cos{(2\omega t + \theta)}}{2\sqrt{(k-4M\omega^2 )^2+4\omega^2\gamma^2}}
\label{Eqn:Prediction}
\end{equation}

Thus, when a voltage is applied to the electrodes, there will be an exponential transient before the free surface displaces by a finite amount determined by the strength of the electric field and the spring's stiffness, with a superimposed oscillation. The system responds at twice the frequency of the driving electric field because the force goes with the square of the voltage. The term $\theta$ is a phase lag caused by the system damping. The constants $A_1$ and $A_2$ are determined by the initial conditions $\delta(0)=\dot{\delta}(0)=0$. The time constants $\tau_1,\tau_2$ satisfy \mbox{ $M\tau^{-2}_{1,2}-\gamma\tau^{-1}_{1,2}+k=0$. }

We consider two limiting cases, DC and high driving frequency. If the applied voltage is constant, the displacement at steady state will be $\delta_{DC}=F_0/k$; \textit{i.e.} the measured force $F=F_0$. If the driving frequency is much greater than the system's natural frequency, this is, $\omega>>\omega_n=\sqrt{k/M}$, the steady state solution will be $\delta_{HF}=F_0/(2k)$, \textit{i.e.} the measured force $F=F_0/2$. 

For the liquids considered in this paper, the system is overdamped, this is, $\gamma>>4Mk$. In this case, the time constants can be approximated by $\tau_1=\gamma/k$, and $\tau_2=M/\gamma$, with $\tau_1$ being the dominating timescale.

\section{Results and Discussion} \label{results}

\subsection{Control 1: surface forces due to electric fields applied across air} \label{Results-air}
\begin{figure}
\centering
\includegraphics{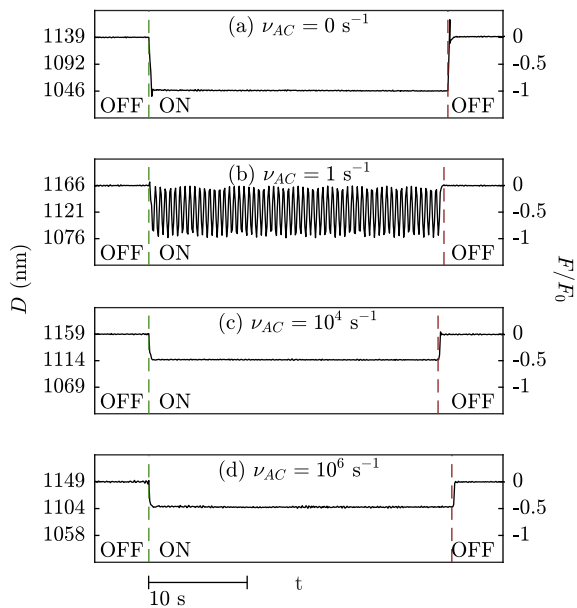}
\caption{Measured displacement when an electric field is applied between the silver surfaces with air as the medium between the mica surfaces. The corresponding force, $F$, is shown normalised by $F_0$ on the right axis. The initial distance for the cases shown is close to 1.15~$\mu$m. For the DC case, the applied voltage was $V_0=$10~V, whereas for the AC cases the applied voltage amplitude was 9.9~V. The spring constant was $k=$110~N/m; mica thickness $T = 3.4~\mu$m;  curvature $R =$ 7.3 mm. }
\label{Figure:Air}
\end{figure}

We begin with control experiments carried out with dry air between the two electrode surfaces. Results of experiments performed with constant and sinusoidal input voltages and at an initial surface separation of approximately 1~$\mu$m are shown in Figure~\ref{Figure:Air}. The left axis gives the absolute surface separation between the electrodes, measured during a period of time when the electric field is off, switched on, and then off. The right axis in Figure~\ref{Figure:Air} shows the measured force, $F$, normalised by $F_0$ for comparison to the predicted values in section~\ref{dielectrics}. 

On application of a DC electric field (\textit{i.e.} $\nu_{\rm{AC}}=0$, Figure~\ref{Figure:Air}(a)) we observe a reduction of the distance $D$ between the surfaces, corresponding to an attractive force between the electrode surfaces. The response (spring deflection) occurs very rapidly in this experiment with air. The attractive force is maintained for the duration that the electric field is on. When the field is switched off the surfaces revert to the original, larger, separation. The magnitude of the measured force is very close to the value of $F_0$ predicted by our simple model (as clear from the right, normalised, axis). 

In Figure~\ref{Figure:Air}(b)-(d) we show the outcomes with AC applied voltages of varying $\nu_{\rm{AC}}$. The natural frequency of the system, estimated from the spring constant and the value of the mass of the lens, is approximately $\omega_n=195$~rad$\cdot$s$^{-1}$ (or $\nu_n =30~\rm{s}^{-1}$) for this particular experiment; thus Figure~\ref{Figure:Air}(b) represents the case $2\pi \nu_{\rm{AC}}<<\omega_n$, whereas Figure~\ref{Figure:Air}(c),(d) are at $2\pi \nu_{\rm{AC}}>>\omega_n$. 

At $\nu_{\rm{AC}} = 1~\rm{s}^{-1}$, well below the natural frequency, the electrode separation oscillates due to the changing magnitude of the surface force during the AC cycles and at twice the frequency of the applied field, as predicted by equation~\ref{Eqn:Prediction}.  At higher frequencies ($10^{2}$ to $10^{6}$ $\rm{s}^{-1}$) there is no oscillatory response but instead a constant displacement, corresponding to a constant force $F=k\delta$ close to our prediction of $F_0/2$. Heuristically, we can say that the electrodes cannot keep up with the oscillating field at these higher frequencies, but instead maintain a separation corresponding to the average force. We performed experiments at $D_0$=1,4 and 16~$\mu$m,  and at frequencies $\nu_{\rm{AC}}=$ 0, 1, $10^2$, 10$^3$, 10$^4$, 10$^5$, and 10$^6$ $\rm{s}^{-1}$, and in all cases the measured spring deflections are within 10$\%$ of the theoretical prediction with no fitting parameters. This small scatter around the theoretical value is likely due to experimental uncertainty in the value of the spring constant ($k=110\pm4.3$ N/m) and the measurement of $R$ from the FECO pattern. 

\subsection{Control 2: surface forces due to electric fields across PDMS oil} \label{Results-pdms}

\begin{figure}
\centering
\includegraphics{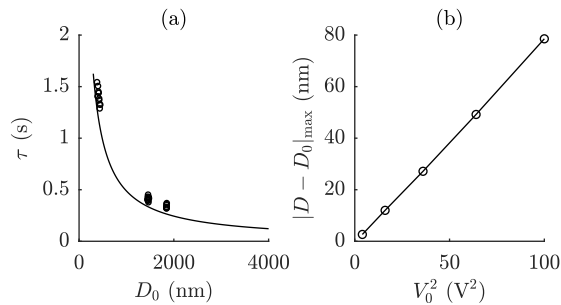}
\caption{Measured displacement (black dots) and calculated trajectory from Equation~\ref{Eqn:Prediction} (solid gold line) when an electric field is applied in the SFB setup with PDMS as the medium between the mica surfaces, at an initial distance of (a) 424 nm and (b) 1471 nm. AC electric fields, with $V_0=7.07$~V, $\nu_{\rm{AC}}=10^{3}$~s$^{-1}$, were applied for 90 seconds. Upon application of an electric field, the gap distance is reduced due to the surface attraction, with a transient timescale determined by the viscous drag (drainage) of the liquid. At $D_0=$424~nm, the measured transition time from an exponential fit to the data is 1.37~s, while $\tau=\gamma/\kappa=$1.15~s. At $D_0=$1471~nm, the transition time extracted from the data is 0.42~s, while $\tau=\gamma/\kappa=$0.33~s. The magnitude of the attractive force due to the AC field is closely similar to the predicted value of $F_0/2$, as seen on the right axis.}
\label{Figure:PDMS}
\end{figure}

The second control experiment is with PDMS oil between the electrodes. The electrostatic force is expected to be of similar form to that observed across dry air (modified in magnitude due to the difference in permittivity), but the mechanical response should now be substantially different due to the viscous force exerted by the liquid. Indeed this is found to be the case: in Figure~\ref{Figure:PDMS} we show typical measurements of the electrode separation vs time during switching on and off of an AC electric field across PDMS. When the field is applied the surfaces move together, with displacement varying exponentially with time towards a steady-state separation. The timescale is $\sim$ 1 s and varies with $D_0$; a thinner liquid film drains more slowly from the confined region between the electrodes. When the voltage is set back to zero, the surface separation relaxes back to the initial value over the same timescale. 

\begin{figure}
\begin{centering}
\includegraphics{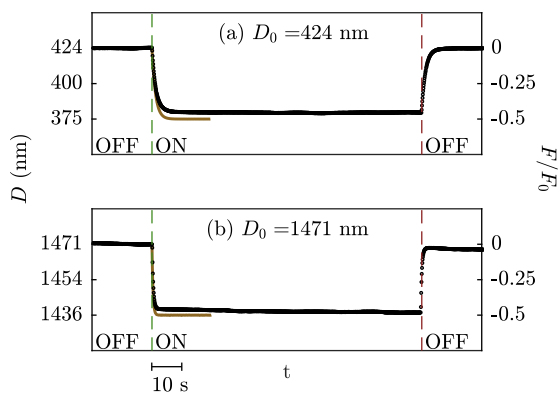}
\caption{(a) Measured relaxation times (dots) in PDMS films for varying applied voltages and frequencies as a function of initial surface separation $D_0$, compared to the model prediction $\tau=\gamma/\kappa$ (solid line). (b) Measured displacement as a function of applied voltage $V_0$, for an initial distance $D_0=3.93~\mu$m and $\nu_{\rm{AC}}=0$~s$^{-1}$. The displacement scales with the square of the applied voltage. }
\label{Figure:PDMS2}
\end{centering}
\end{figure}

The mechanical response under the electrostatic and viscous forces is well described by Equations~\ref{Eqn:Prediction} and ~\ref{Eqn:F0vsV}. This is seen in Figure~\ref{Figure:PDMS} where a single trajectory is compared to Equation~\ref{Eqn:Prediction} (no fitting parameters), and more directly in Figure~\ref{Figure:PDMS2} where we compare in (a) the timescale of the measured response to the calculation of Equation~\ref{Eqn:Prediction} (no fitting parameters) and in (b) the magnitude of the deflection (which is proportional to the attractive force) with the square of the voltage. Again, as in the control experiment with air, we find that the magnitude of the displacement is within 10$\%$ of the model over the range of initial distances studied ($D_0=0.4, 1.4, 1.9, 4$ and 16~$\mu$m) and at all frequencies studied. 

\subsection{Surface forces due to AC fields across ionic liquids} \label{Results-IL}
\begin{figure}
\begin{centering}
\includegraphics{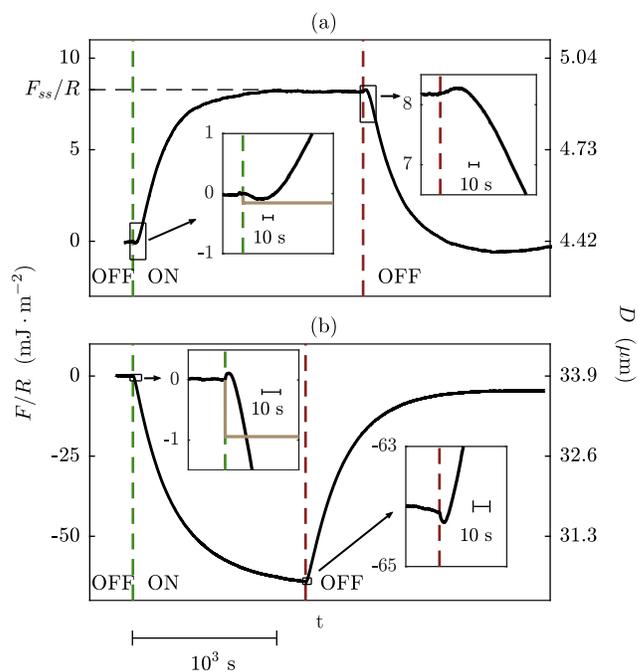}
\caption{Measured force (normalised by radius) between mica surfaces across [C$_2$C$_1$Im][NTf$_2$] ionic liquid as a function of time during switching on and off the electric field. (a) Example showing the slow evolution towards a repulsive steady-state in AC electric field. The force at steady state is denoted $F_{ss}$, as marked by the horizontal dashed line. The insets show the detail of the response immediately after the field is turned on and off. When the field is turned on (indicated by the first vertical green dashed line), there is an immediate attraction over a few seconds (see first inset), which is then followed by a strong repulsion over a longer timescale. The calculated force for a (hypothetical) dielectric liquid of equivalent dielectric constant and viscosity is plotted for comparison with the solid golden line in the inset.  When the field is turned off (indicated by the vertical red dashed line) there is a quick repulsion over a few seconds (second inset), followed by a slow relaxation until the system returns to its original state. For this example, $D_0=4.4~\mu$m, $V_0=2.82$~V, $\nu_{\rm{AC}}=10^4$~s$^{-1}$, $T=2.8525~\mu$m, $k=150$~N/m$^2$, and $R=9.3$~mm. (b) Example showing the slow evolution towards an attractive steady-state in AC electric field. The insets show the details of the response immediately after the field is turned on and off. When the field is turned on, there is an immediate repulsion over a few seconds, which is then followed by the strong attractive force. Again the calculated force for a hypothetical dielectric liquid with the same viscosity is plotted as a solid golden line in the first inset. When the field is turned off there is a quick attraction over a few seconds (second inset), followed by a slow relaxation until the system returns to its original state. For this example, $D_0=33.9~\mu$m, $V_0=7.07$~V, $\nu_{\rm{AC}}=~10^5$~s$^{-1}$, $T=2.95~\mu$m, $k=179$~N/m, and $R=9.6$~mm.  }
\label{Figure:IL1}
\end{centering}
\end{figure}

With the control experiments in hand and the mechanics of the experiment well characterised, we now report our measurements with electrolyte (ionic liquid) between the electrodes. Example measurements are shown in Figure~\ref{Figure:IL1}, where the setup includes mica spacers, and in Figure~\ref{Figure:AuILAu}, where the electrolyte is directly contacting the gold electrodes with no mica spacers (\textit{i.e.} $T=0$). The forces measured between the electrodes across ionic liquids with oscillating electric field are very substantially different -- in timescale, force magnitude, and other features --  compared to the simple dielectric examples.

\subsubsection*{General form of the measured force} 
\begin{figure*}
\centering
\includegraphics[scale=1.0]{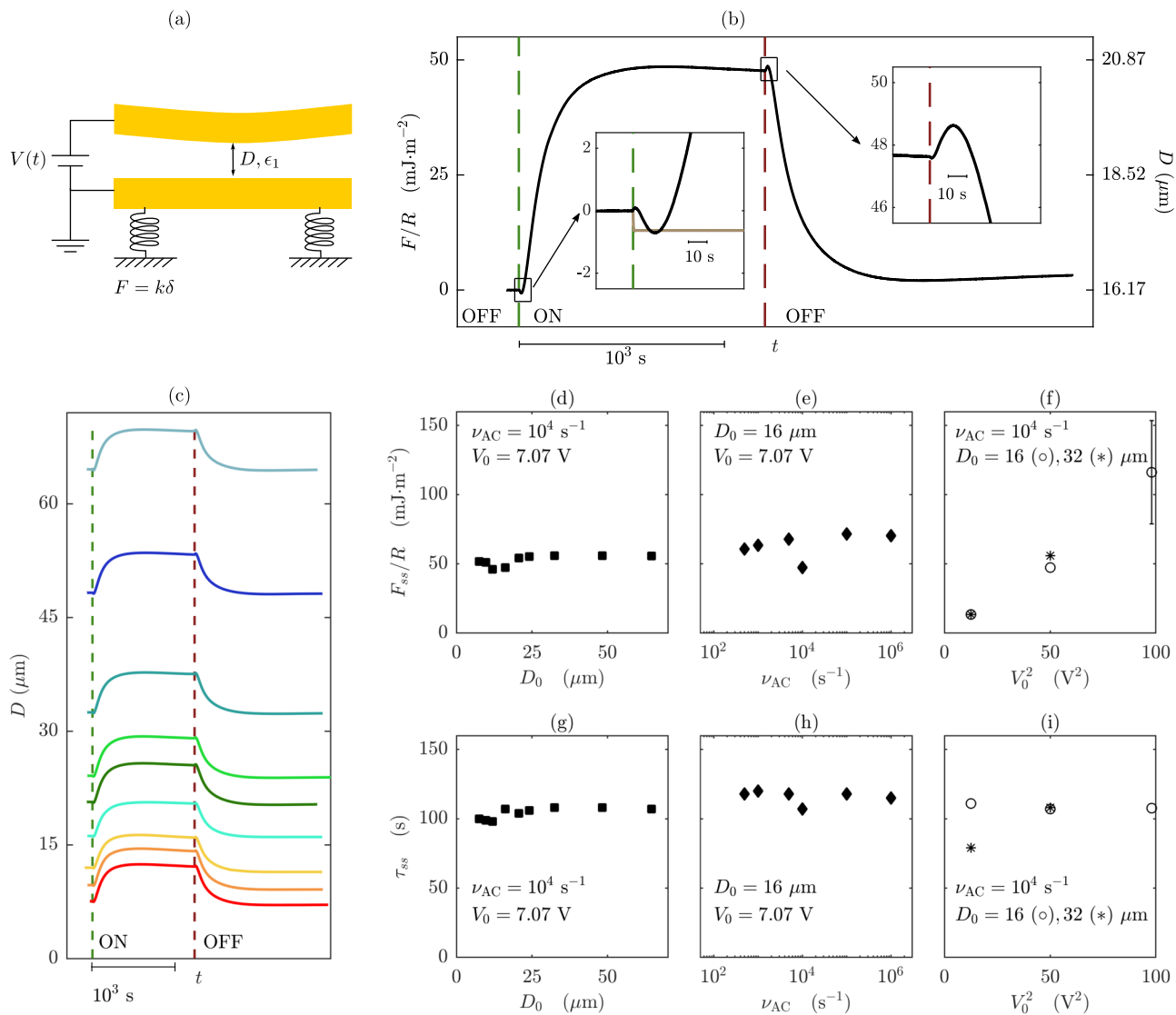}

\caption{ Measurements of the force between electrode surfaces across an ionic liquid, [C$_4$C$_1$Pyrr][NTf$_2$], during switching on and off of an AC electric field. Measurements were made in the SFB with no mica spacer layers, instead the liquid is in direct contact with the gold electrodes. 
(a) Schematic of the SFB setup with Au electrodes and no mica spacers ($T=0$) 
(b) Force (normalised by radius of  curvature) as a function of time. All results in this figure are for a single set of gold surfaces; for this experiment, the steady state force was repulsive. The field was switched on  at the point marked with green vertical dashed line, and off at the red vertical dashed line. An overall repulsive force is observed when an AC electric field is applied; the insets show the detail of the response immediately after the field is turned on and off. When the field is turned on, there is a repulsive interaction over $\sim$ 1 s, followed by an attraction over a few seconds, before the force becomes strongly repulsive. The calculated response for an equivalent dielectric liquid (of the same viscosity and dielectric constant), is plotted as a solid golden line in the inset. Once the force becomes repulsive, it takes several hundred seconds for the system to achieve steady state. When the field is turned off there is a quick attraction, followed by a repulsive force over a few seconds, before the force relaxes back towards zero over $\sim 10^3$~s. For this example, $V_0=7.07$~V, $D_0=16.17~\mu$m, and $\nu_{\rm{AC}}=10^4$~s$^{-1}$.
(c) Traces showing the variation in electrode separation with time measured for a range of initial separation distances (all measured at $\nu_{\rm{AC}}=10^4$~s$^{-1}$, $V_0$=7.07~V). 
Panels~(d),~(e)~and~(f) show the dependence of $F_{ss}$ on initial distance $D_0$, driving frequency $\nu_{\rm{AC}}$, and voltage amplitude $V_0$.  Panels (g), (h) and (i) show the dependence of the experimentally measured asymptotic time to reach steady state, $\tau_{ss}$ as a function of initial distance $D_0$, driving frequency $\nu_{\rm{AC}}$, and voltage amplitude $V_0$. In panels (d) and (g), $\nu_{\rm{AC}}$ and $V_0$ were kept constant at $10^4$~s$^{-1}$and 7.07~V, respectively, while varying the surface separation from 8~to~64~$\mu$m. In panels (e) and (h), the initial distance $D_0$ was kept at 16~$\mu$m and the voltage amplitude was $V_0$=7.07~V, while varying the electric field frequency from 500~s$^{-1}$ to $10^6$~s$^{-1}$. In panels (f) and (i), the frequency was kept at $\nu_{\rm{AC}}=10^4$~s$^{-1}$.  The open circles represent measurements performed for an initial surface separation of approximately $D_0\sim16~\mu$m, while the asterisks represent measurements from $D_0\sim32~\mu$m. For all the measurements reported in this figure, $T=0~\mu$m, $k=150$~N/m, and $R=1.41$~mm. The trends reported in this figure and the text, namely, that $F_{ss}$ scales with $V_0^2$ but is independent of $D_0$ and $\nu_{\rm{AC}}$, and that $\tau_{SS}$ is independent of the applied frequency, voltage or electrode separation, were consistent across all our experiments including those with mica spacers, and regardless of the sign of the steady state force (attraction or repulsion).}


\label{Figure:AuILAu}
\end{figure*}

 In Figure~\ref{Figure:IL1} we show examples of the measured surface force as a function of time during application of a sinusoidally oscillating potential difference between the electrodes. The trace shows the force to be zero before application of the field (region marked `OFF'), then, when the field is applied (marked `ON'), the force evolves over time and eventually reaches a steady-state after $\sim 10^3 \rm{s}$. We denote the force between the surfaces at steady-state in the applied AC field as $F_{ss}$, as marked in Figure~\ref{Figure:IL1}(a). When the field is then switched off, the force reverts towards zero over a similar timescale and with similar (but inverted) features. In some experiments the force at steady state in the AC field was attractive, and in others it was repulsive; in Figure~\ref{Figure:IL1} 
we show one example of each type of behaviour. The time-evolution of the force after switching on the field is quite complex: the direction of the force changes during each single trajectory from attractive to repulsive (as in the insets to the top trace) or from repulsive to attractive (insets to the bottom trace). Despite these complex features, the behaviour was found to be highly reproducible and reversible. Furthermore, we observed very similar behaviour with and without the mica spacer layers: an example from a measurement with no mica layers is shown in Figure~\ref{Figure:AuILAu}.  The dependence on $V_0$, $\nu_{\rm{AC}}$, $D_0$ has been measured and we outline some key features below. 

\subsubsection*{Onset frequency: `beating the double-layer' and a measure of the double-layer timescale} 
Forces such as those in Figure~\ref{Figure:IL1} 
are observed only at sufficiently high AC frequency. At low $\nu_{\rm{AC}}$ the forces tend towards the DC limit, where screening electrical double layers form and the field does not penetrate into the electrolyte: at $\nu_{\rm{AC}}=0$  and $D>>\lambda_S$, $F_e = 0$ at equilibrium due to exponential screening of the electric field (see Equation~\ref{PiEDL}). 
Only when the field oscillates more rapidly than the timescale for double layer formation will any influence of the field be felt within the body of the electrolyte at steady state:  heuristically, if the electric field applied across an electrolyte film varies with time sufficiently rapidly that ions cannot `keep up' with the field -- \textit{i.e.} cannot translate over the distances required to form a double layer during one cycle --  then the field will be unscreened.
 In our experiments the onset of long-range interactions -- such as those in Figure~\ref{Figure:IL1} 
(at $D \sim 10~\mu \rm{m} >>\lambda_S)$ -- occurs at AC frequencies $\nu_{\rm{AC}}^{\rm{onset}} \sim \mathcal{O}(10^2~\rm{s^{-1})}$, i.e. the timescale of double-layer formation, $\tau_{DL} \sim \mathcal{O}(10^{-2} \rm{s})$. This empirical screening timescale is consistent with the theoretical double-layer charging time, $\tau_{c}$ for parallel plate electrodes\cite{Bazant2004}: 
\begin{equation}
\tau_{c} \sim \frac{\lambda_S D}{\mathcal{D_{+/-}}}
\label{Eqn:tauDL}
\end{equation}
where $\mathcal{D_{+/-}}$ is the mean ion self-diffusion coefficient. For the ionic liquid used here $\lambda_S = 8 ~\rm{nm}$\cite{Lee2017}, $\mathcal{D_{+/-}} \approx 5 \times 10^{-11} ~\rm{m^2 s^{-1}}$\cite{Tokuda2006}, yielding $\tau_{c} \approx 0.2 \times 10^{-2} \rm{s}$. We can read from this agreement in order of magnitude between these measured and predicted double layer timescales that the formation of the double layer in response to perturbation of the electrode voltage involves migration over distances comparable to the slit width, $D$, as well as reorganisation within the layer of thickness $\lambda_S$. It is expected that smaller slit widths will lead to altered scaling of  $\tau_{DL}$: a recent experimental study\cite{Tivony2018} of $\tau_{DL}$ with dilute electrolytes and smaller $D$ indicated that the transmission line model\cite{DeLevie1963,Mirzadeh2014} - incorporating a geometrical factor to account for the pore dimensions - was a good representation of the measured timescales.  \\

\subsubsection*{Force magnitude and timescale for reaching steady state in an AC field} 
The magnitude of the force at steady state, and the timescale for evolution towards steady state after imposing the electric field, are both substantially larger than would be the case for a (hypothetical) dielectric liquid with the same viscosity and zero-frequency dielectric constant; this can be clearly seen by comparing the measured and calculated trajectories in Figures~\ref{Figure:IL1} and \ref{Figure:AuILAu}  (the calculations are shown as solid gold lines in the insets in both Figure~\ref{Figure:IL1} and Figure~\ref{Figure:AuILAu}).

In order to test directly the effects of electrode separation distance, field magnitude, and frequency on the measured force, we carried out a series of experiments with no (mica) spacer layers between the electrodes and the liquid (\textit{i.e.} $T=0$); Figure~\ref{Figure:AuILAu}. In this case all of the voltage drop occurs across the liquid and so variation of the force with potential difference can be interpreted unambiguously. A diagram of this setup and an example trace of the force \textit{vs.} time are shown in Figure~\ref{Figure:AuILAu}(a) and (b). In Figure~\ref{Figure:AuILAu}(c), we show traces recorded at a range of electrode separation distances, whilst holding other parameters constant. It is clear that the force at steady state, $F_{ss}$, and the timescale for reaching steady state are similar in each trace despite the wide variation of $D_0$. In order to quantify the timescale of reaching eventual steady-state in traces such as this, we fitted the asymptotic shape (i.e. disregarding the small oscillations in direction at short times) to a single exponential function, 
 \begin{equation}
 F = F_{ss} \left(1-e^{-t/{\tau_{ss}}}\right),
 \end{equation}
 with the fitting parameter $\tau_{ss}$. An example of the closeness of fit is shown in the Supplementary Information.  By carrying out series of measurements such as that in Figure~\ref{Figure:AuILAu}(c) -- varying one parameter at a time within a single experiment -- we checked the influence of $D_0$, $\nu_{\rm{AC}}$, and $V$ on $F_{ss}$ and $\tau_{ss}$: examples from a single series of measurements are plotted in Figure~\ref{Figure:AuILAu}(d)-(i). It is clear that there is no significant influence of $D_0$ or $\nu_{AC}$ on $F_{ss}$, whilst $F_{ss} \sim V^2$ over the range studied. These trends were found to be consistent across all our experiments, even though some experiments showed overall attractive and others overall repulsive behaviour. The scaling of $F_{ss}$ with $V^2$ is reminiscent of the behaviour of dielectrics (Equation~\ref{Eqn:F0vsV}) and also as anticipated by recent work on electric field effects on electrolytes by Hashemi Amrei \textit{et al.}\cite{Hashemi2018}. 

The timescale for evolution of the force to the steady-state value, $\tau_{ss}$, is of the order of $10^2~\rm{s}$ and so is substantially slower than the viscous timescale calculated for fluid drainage away from the film under a steady force. The timescale appears to be independent of $D_0$, $\nu_{AC}$, and $V$. The origin of this slow evolution of the force, arising from the action of the electric field (and possibly also its lateral gradient) on the ions in the bulk, is not yet clear. However, we note that several theoretical analyses predict that the dynamics of the electrolyte after a suddenly imposed DC field relax exponentially with more than one characteristic timescale: the faster is the charging timescale, $\tau_c$, (as discussed above) and the slower timescale corresponding to the diffusive timescale in the geometry of the experiment, $\tau_{\mathcal{D}}$\cite{Bazant2004,Khair2018}. In those theoretical works, the system geometry is infinite in the directions perpendicular to the field so the relevant lengthscale determining $\tau_{\mathcal{D}}$ is necessarily the electrode-electrode separation, $D$, with $\tau_{\mathcal{D}}$ scaling as $\tau_{\mathcal{D}} \sim D^2/\mathcal{D}$ (with $\mathcal{D}$ a composite diffusivity to be discussed shortly) \cite{Khair2018}. It was also noted that, with curved surfaces (such as between colloidal particles) the diffusive timescale can depend on the particle diameter\cite{Bazant2004}; this might imply a timescale $\sim R\cdot D/\mathcal{D}$ to account for the curvature in our experiments. However, in our experiments we observed that over the range $D_0 =~ 8~\mu{\rm m}$ to $D_0= ~65~\mu{\rm m}$ (all with $T=0$)  $\tau_{\rm{ss}}$ is independent of $D_0$; it appears that $ \tau_{\rm{ss}}$ is entirely determined by lengthscales other than the film thickness (at least in the case where $R \gg D \gg \lambda_S$, as in all the present measurements). Also interesting is the question as to what exactly is the relevant diffusivity determining the diffusive timescale. Balu and Khair's analysis of the coupling of ion fluxes in response to an imposed field concludes that $\tau_{\mathcal{D}}$ scales inversely with the \textit{ambipolar diffusivity} of the salt, $\mathcal{D}_a$, quantifying the difference in diffusivity of the anions and cations with respect to solvent\cite{Khair2018}. In the present experiments we study just two (solvent-free) ionic liquid electrolytes and so it is not yet possible to make comparison to this quantity; it will be important to uncover in future the relationship between $\mathcal{D}_a$, $R$, and $\tau_{\rm{ss}}$.

\subsubsection*{Bifurcation in the direction of the force} 

Quite remarkable is the observation that -- in different experiments -- the direction of the force at steady state was sometime attractive and other times repulsive (as in Figure~\ref{Figure:IL1}(a) and (b); also see the SI for a summary).  Within each experiment (same liquid, same electrodes etc.), the behaviour was entirely reproducible, and most notably the features apparent in the transient are common (but inverted) between experiments showing attraction and those showing repulsion at steady state. We found no general correlation between frequency, electrode material or voltage and the direction of the force. In addition, between the two ionic liquids studied there was not a qualitative or clear quantitative difference in the response, and with each ionic liquid both attractive and repulsive behaviour has been observed. Whilst considering the origin of the measured forces, it is important to consider the possibility of either Joule heating or electrochemical (Faradic) irreversible processes playing a role; we have considered these possible artifacts in detail and explain in the Supplementary Information our reasons for having discounted them.  We note the strong connection between our observation of this bifurcation to (i) measurements of colloidal forces under AC fields in electrolyte, in which both attraction and repulsion have been reported\cite{Woehl2015,Bukosky2015}, and (ii) the theoretical report of non-linear AREF (Asymmetric Rectified Electric Field) effects where it was pointed out that oscillating electric fields in liquids (electrolytes) create a long-range steady field with the absolute direction of the steady field can switch depending on the starting distance and AC frequency\cite{Hashemi2018}. A direct AREF field could act to generate a force dependent on the asymmetry of the ion mobilities. In addition, we note that in our experimental geometry with curved electrode surfaces, the electric field has a lateral gradient away from the centre and this may also drive ion flux in the direction perpendicular to the field. To probe these hypotheses future investigations must involve a range of electrolytes, including diluted electrolytes and ionic liquids with asymmetry in cation and anion diffusivity, and variation of electrode curvature or geometry.


\section{Summary}
We have demonstrated that the direct measurement of surface forces between electrodes can be a sensitive tool for determining the electrostatic and dynamic properties of fluids under the influence of an electric field. Control experiments with air and a viscous dielectric liquid allowed us to fully characterise the measured forces in the absence of free charges in the medium: the equation of motion, incorporating the viscous drag of the fluid, was solved and used to fully describe the measured trajectory with no fitting parameters. 

In contrast, our direct measurements of the surface force when an oscillating electric field is applied across an electrolyte (ionic liquid) revealed forces of substantially larger magnitude, and with slow evolution towards steady state, than anticipated for a similar fluid without free charges. Such substantial forces acting across electrolytes are likely to be important for many practical phenomena ranging from energy storage to colloidal systems, and imply new routes to controlling surface interactions such as friction and adhesion. We note the connection between our measurements and several reports of electric field effects across electrolytes in the literature, however the details of their origin is not yet clear and more experimental evidence is required. 

The technique and analysis reported here are very general and could, in future, be applied to any systems of soft or biological matter as well as liquid crystals, electrolytes, polymers, \textit{etc.} and opens the way to investigation of the complex non-equilibrium phenomena occurring in soft systems subjected to externally applied fields.


\section*{Conflicts of interest}
There are no conflicts to declare.

\section*{Acknowledgements}
The authors are particularly grateful to Carlos Drummond for discussions and sharing of findings on this topic over many years; his related measurements with diluted electrolytes are to be reported separately. We are grateful to co-workers Alexander Smith, Florian Hausen, Nico Cousens, Christian van Engers, and Marco Balabajew who carried out experimental work which motivated and contributed to this project. James Hallett and Marco Balabajew are thanked for sharing codes for processing some of the data, and Joelle Frechette for providing the PDMS. We also acknowledge helpful discussions with Romain Lhermerout, Alpha Lee, Yan Levin, Ramin Golestanian and Bruno Zappone. S.P. and C.S.P-M. are grateful for funding from The Leverhulme Trust (RPG-2015-328) and the European Research Council (under Starting Grant No. 676861, LIQUISWITCH).  



\balance


\bibliography{EFields.bib} 
\bibliographystyle{rsc} 

\end{document}